\documentclass[utf8]{frontiersSCNS} 

\usepackage{url,hyperref,lineno,microtype,subcaption}
\usepackage[onehalfspacing]{setspace}
\usepackage{caption}
\usepackage{threeparttable}

\def\keyFont{\fontsize{8}{11}\helveticabold }
\def\firstAuthorLast{Xue {et~al.}} 
\def\Authors{Jianchao Xue\,$^{1}$, Hui Li\,$^{1,2,*}$ and Yang Su\,$^{1,2}$}
\def\Address{$^{1}$Key Laboratory of Dark Matter and Space Astronomy, Purple Mountain Observatory, Chinese Academy of Sciences, 10 Yuanhua Road, Nanjing 210023, People’s Republic of China \\
$^{2}$School of Astronomy and Space Science, University of Science and Technology of China, 96 Jinzhai Road, Hefei 230026, People’s Republic of China}
\def\corrAuthor{Hui Li}

\def\corrEmail{nj.lihui@pmo.ac.cn}

\begin{document}
\onecolumn
\firstpage{1}

\title[Spectral evolution of an eruptive prominence]{Spectral evolution of an eruptive polar crown prominence with IRIS observations} 

\author[\firstAuthorLast ]{\Authors} 
\address{} 
\correspondance{} 

\extraAuth{}

\maketitle

\begin{abstract}

\section{}
Prominence eruption is closely related to coronal mass ejections and is an important topic in solar physics. Spectroscopic observation is an effective way to explore the plasma properties, but the spectral observations of eruptive prominences are rare. In this paper we will introduce an eruptive polar crown prominence with spectral observations from the Interface Region Imaging Spectrograph (IRIS), and try to explain some phenomena that are rarely reported in previous works. The eruptive prominence experiences a slow-rise and fast-rise phase, while the line-of-sight motions of the prominence plasma could be divided into three periods: two hours before the fast-rise phase, opposite Doppler shifts are found at the two sides of the prominence axis; then, red shifts dominate the prominence gradually; in the fast-rise phase, the prominence gets to be blue-shifted. During the second period, a faint component appears in Mg~\textsc{ii} k window with a narrow line width and a large red shift. A faint region is also found in AIA $304\,\mathrm{\AA}$ images along the prominence spine, and the faint region gets darker during the expansion of the spine. We propose that the opposite Doppler shifts in the first period are a feature of the polar crown prominence that we studied. The red shifts in the second period are possibly due to mass drainage during the elevation of the prominence spine, which could accelerate the eruption in return. The blue shifts in the third period are due to that the prominence erupts toward the observer. We suggest that the faint component appears due to the decreasing of the plasma density, and the latter results from the expansion of the prominence spine.
\tiny
 \keyFont{ \section{Keywords:} Sun: corona -- Sun: coronal mass ejections (CMEs) -- Sun: filaments, prominences -- Sun: UV radiation -- techniques: spectroscopic} 
\end{abstract}

\section{Introduction}

Solar prominences are composed of cold and dense plasma suspended in the hot corona \citep{2010SSRv..151..243L,2014LRSP...11....1P,2015ASSL..415.....V,2020RAA....20..166C}. Prominence eruptions have a close relationship with flares and coronal mass ejections (CMEs), and the latter two phenomena are main causes of the space weather storms. Hence studying the triggering mechanism and evolution of prominence eruptions are important topics in solar physics. Spectroscopic observation is an effective way to reveal plasma properties and line-of-sight (LOS) motions. However, high-quality spectral data of eruptive prominences are rare due to limited field of view (FOV) of general spectroscopic observations and randomness of prominence eruptions.

The Interface Region Imaging Spectrograph \citep[IRIS, ][]{2014SoPh..289.2733D} is a small explorer spacecraft launched in 2013 June. It provides simultaneous high-resolution spectral and imaging data from the photosphere to the corona. The IRIS especially has an advantage of observing chromosphere and transition region with some strong resonance lines of Mg~\textsc{ii} (temperature of formation of $\mathrm{log}T\mathrm{[K]}\sim 4.0$), C~\textsc{ii} ($\mathrm{log}T\mathrm{[K]}\sim 4.3$), and Si~\textsc{iv} ($\mathrm{log}T\mathrm{[K]}\sim 4.8$). The prominence core has a chromospheric temperature, and prominence also has a prominence-corona transition region (PCTR). So the IRIS is also suitable to observe prominences and filaments. The IRIS has been widely used to study the dynamics of quiescent prominences \citep{2014A&A...569A..85S,2016SoPh..291...67V,2016ApJ...831..126O,2018ApJ...865..123R}, but the spectroscopic observations of eruptive prominences are still rare. Among the few works, \citet{2015ApJ...803...85L} reported an erupting prominence in active region using IRIS observations; the authors found a faint component with a LOS velocity up to $460\,\mathrm{km\,s^{-1}}$, and revealed the unwinding motions during the prominence eruption. \citet{2019A&A...624A..72Z} studied an eruptive prominence in quiet region with radiative transfer computations; they derived the electron densities of the prominence between $1.3\times 10^9$ and $6.0\times 10^{10}\,\mathrm{cm^{-3}}$, the mean temperature around $1.1\times 10^4\,\mathrm{K}$, and the total hydrogen mass between $1.3\times10^{14}$ and $3.2\times 10^{14}\,\mathrm{g}$. 

In this work, we focus on the spectral evolution of an eruptive polar crown prominence (the prominence located at high latitude) on 2015 April 28th, which erupts successfully with a CME. The high-quality IRIS observations reveal some phenomena that have not been reported, and we try to give reasonable explanations on them. This event was studied by \citet{2021ApJ...906...62L} using extreme ultraviolet (EUV) images, who were interested in the outflows within the dimming region and proposed that the outflows are the origin of CME-induced solar wind. Dai et al. (submitted) used EUV images to explore the eruption mechanism of this event, and thought that the eruption is related to the prominence oscillation and mass drainage. Our paper is organized as follows: section~\ref{sec:met} introduces the observations and data reduction; section~\ref{sec:res} shows the prominence eruption process and its spectral features; we give our explanations on some observed phenomena in section~\ref{sec:dis}, which is followed by conclusion in section~\ref{sec:con}.

\section{Observations and Data Reduction} \label{sec:met}
\begin{figure}[h!]
    \begin{center}
    \includegraphics[width=7in]{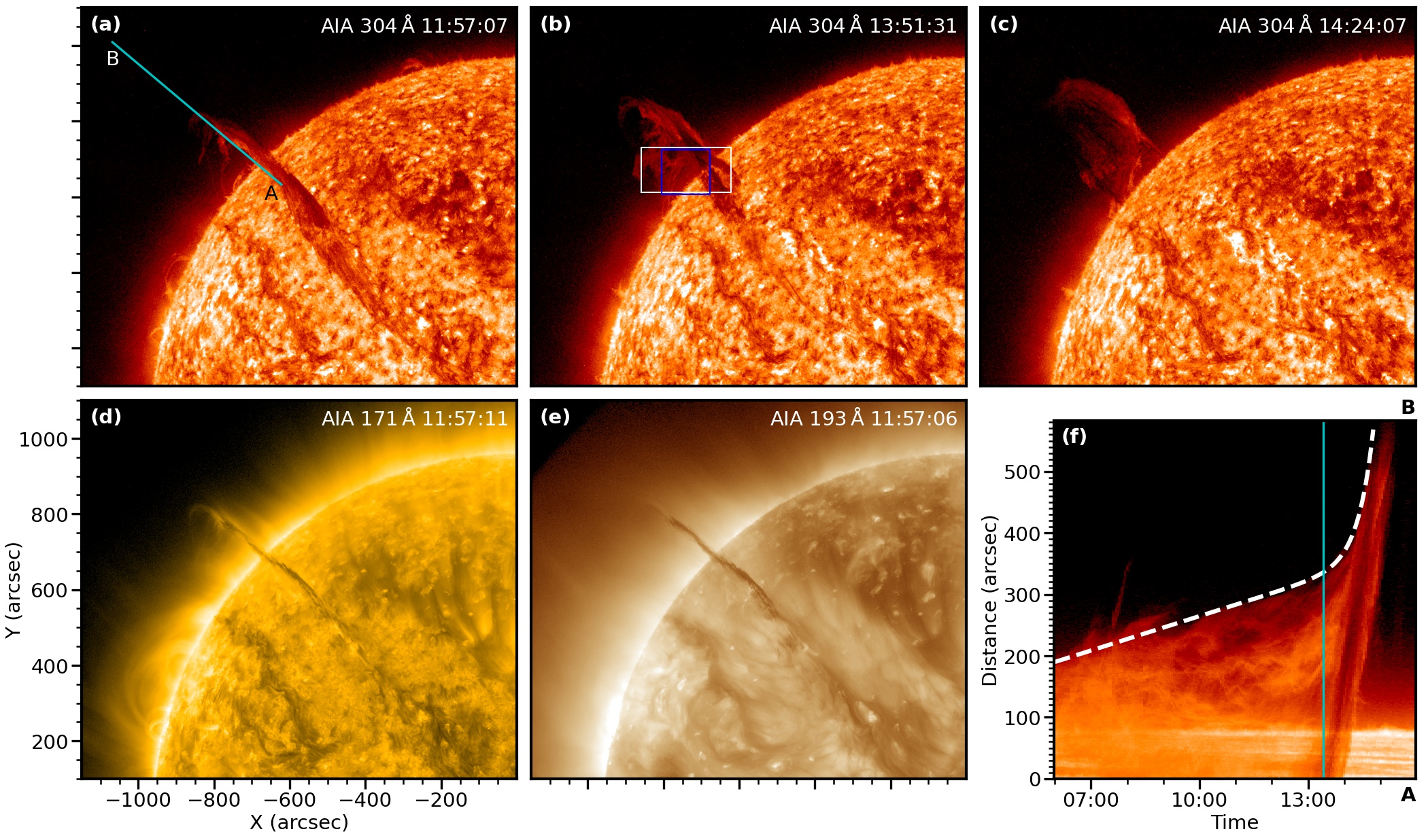}
    \end{center}
    \caption{\label{fig:aia_evo} Evolution of the eruptive prominence in EUV images. (a)-(c) AIA $304\,\mathrm{\AA}$ images. The cyan slice AB in (a) is the position of the time-distance diagram in panel (f). The white box in (b) represents the FOV of IRIS SJI observations, and the blue box marks the FOV of spectroscopic observations. (d)-(e) AIA 171 and $193\,\mathrm{\AA}$ images, respectively. (f) Time-distance diagram along the slice AB in panel (a) from AIA $304\,\mathrm{\AA}$ images. The cyan line marks the onset of the fast-rise phase.}
    \end{figure}

An eruption of a polar crown prominence on 2015 April 28th was well observed by the IRIS. Figures~\ref{fig:aia_evo}(a)-(e) show the snapshots of the eruptive prominence in EUV images from the Atmospheric Imaging Assembly \citep[AIA, ][]{2012SoPh..275...17L} onboard the Solar Dynamics Observatory \citep[SDO, ][]{2012SoPh..275....3P}. The observational channels and times are denoted in each panel. The IRIS observations were carried out between 10:59--14:54~UT with 24 large coarse raster scans, and each scan had $64\times 2''$ raster steps. The binned pixel size along the raster slit is $0.33''$, the same as the slit width. The resulting FOV is $126''\times 119''$ centered at $-742''$ east and $666''$ north (see the blue rectangle in Figure~\ref{fig:aia_evo}(b)). In the snapshot of a slit-jaw image (SJI) at $2796\,\mathrm{\AA}$ in Figure~\ref{fig:iris_cog}(a), the dashed white lines mark the slit position No.30 (left), upper boundary of the FOV of the Mg~\textsc{ii} raster (top), and the slit position No.64 (right). The exposure time of each raster is 8~s and the step cadence is 9.2~s. The spectral resolution is $\sim 5.5\,\mathrm{km\,s^{-1}}$ with binned pixels. The slit occurs at the center of SJIs in Solar-X direction. The FOV of each SJI is $117''\times 119''$, and FOV of SJI observations is marked in Figure~\ref{fig:aia_evo}(b) with the white rectangle. The SJIs at 2796, 1400, and $1330\,\mathrm{\AA}$ are available, which have a cadence of $\sim 37\,\mathrm{s}$ and binned pixel size of 0.333 arcsec in each channel. We mainly use SJIs $2796\,\mathrm{\AA}$ and $1400\,\mathrm{\AA}$, the former has a passband of $4\,\mathrm{\AA}$ centered at $2796\,\mathrm{\AA}$, mainly contributed by the Mg~\textsc{ii} k line; the latter has a passband of $55\,\mathrm{\AA}$ centered at $1390\,\mathrm{\AA}$, mainly contributed by Si~\textsc{iv} lines but also including O~\textsc{iv} lines, etc. 

IRIS level 2 data are used, for which dark current and offsets are removed, flat field is corrected, and geometric and wavelength calibrations (for spectrograph channels) are done. The FOV of SJIs is checked by comparing them with the AIA $304\,\mathrm{\AA}$ images. The spatial position of spectra is shifted along Solar-Y slightly using the fiducial marks on the slit (see Figure~\ref{fig:iris_cog}(b) at $618''$ and $708''$, respectively). The wavelength calibrations are checked using lines Ni~\textsc{i} $2799.474\,\mathrm{\AA}$ in the Mg~\textsc{ii} window and Fe~\textsc{ii} $1392.82\,\mathrm{\AA}$ in the Si~\textsc{iv} window radiated from the solar disk; the wavelength errors are expected to be within $2\,\mathrm{km\,s^{-1}}$. For images of the Si~\textsc{iv} $1394\,\mathrm{\AA}$ line spectra, bright and isolated pixels are identified as spikes, and their values are replaced by their surrounding mean values. Errors of spectral intensities from signal uncertainty and readout noise are considered. The former is set to be square root of photon number, and the conversion coefficient from digital number (DN) to photons is 18 for near ultraviolet (NUV, including the Mg~\textsc{ii} k line), and 4 for far ultraviolet (FUV, including the Si~\textsc{iv} $1394\,\mathrm{\AA}$ line). The readout noise is related to dark current uncertainty, which is set to be 1.8 DN for NUV (from the negative values in data) and 3.3 DN for FUV \citep{2014SoPh..289.2733D}.

EUV images from the SDO/AIA and LOS magnetograms from the SDO/HMI \citep[Helioseismic and Magnetic Imager, ][]{2012SoPh..275..207S} are used. The former has pixel size of $0.6''$ and temporal resolution of 12~s; the latter has pixel size of $0.5''$. Both AIA and HMI images are processed to level 1.5.

\subsection{Estimation of prominence LOS velocity} \label{sec:cog}
\begin{figure}[h!]
    \begin{center}
    \includegraphics[width=7in]{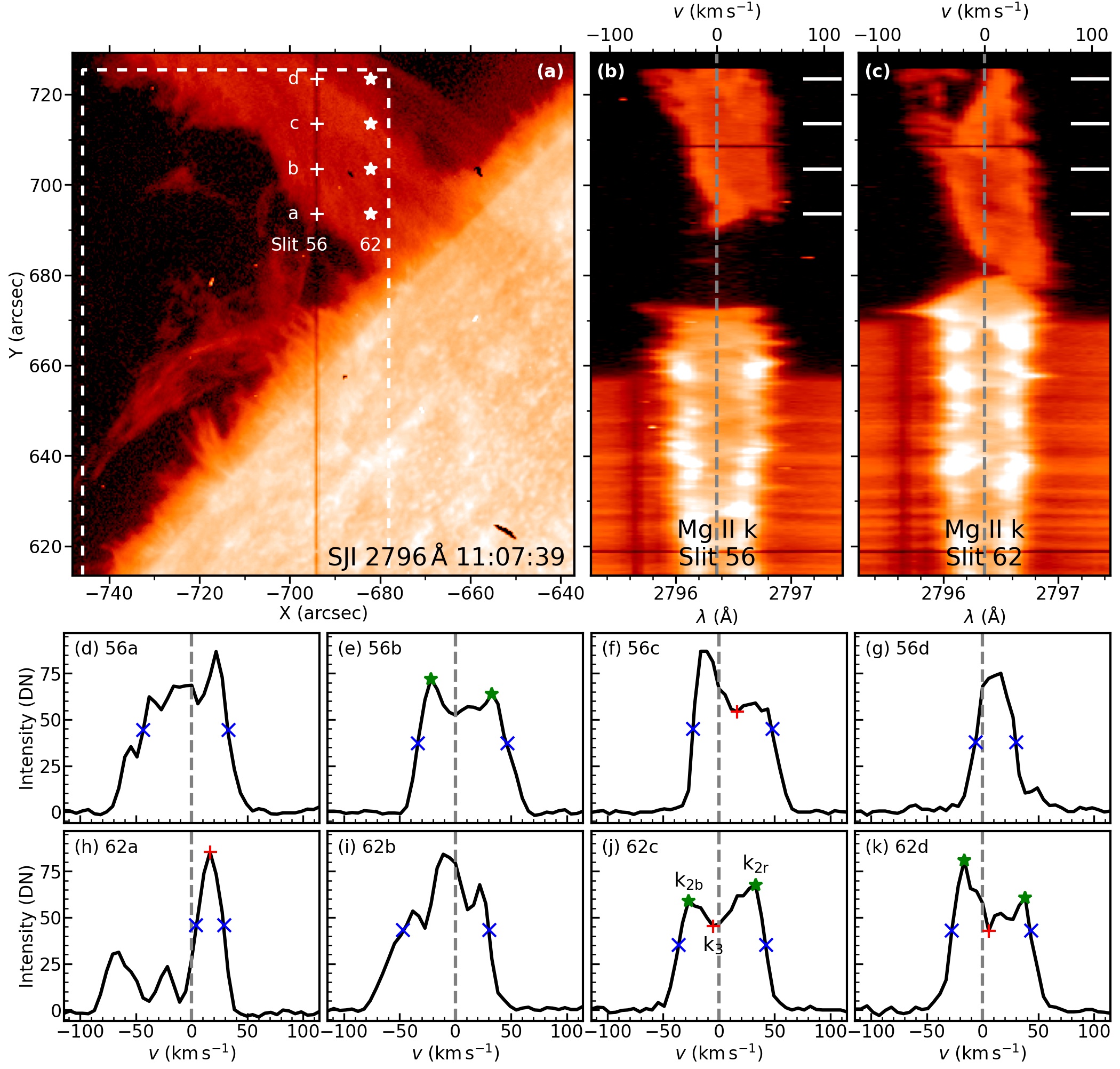}
    \end{center}
    \caption{\label{fig:iris_cog} Estimation of LOS velocities from Mg~\textsc{ii} k line profiles. (a) SJI $2796\,\mathrm{\AA}$. The dashed white lines mark the slit position No.30 (left), upper boundary of the FOV of Mg~\textsc{ii} raster (top), and the slit position No.64 (right). Eight points marked by plus or star symbols are chosen for analysis in panels (d)-(k). (b)-(c) Images of the Mg~\textsc{ii} k spectra along the slit positions No.56 (at 11:07:39~UT) and 62 (at 11:08:44~UT), respectively. The short white lines mark the positions of the spectra in panels (d)-(k). The vertical dashed lines represent the rest wavelength of the Mg~\textsc{ii} k line. (d)-(k) Mg~\textsc{ii} k profiles at the positions marked in panels (a)-(c). The red pluses represent the positions of $\mathrm{k_3}$, the green stars represent $\mathrm{k_2}$, and the blue $\times$ symbols represent the positions of half maximum.}
\end{figure}   

It is almost impossible to describe LOS (Doppler) velocities of a prominence strictly, due to that a prominence consists of many threads with different velocities \citep{2014A&A...569A..85S,2016ApJ...831..126O}. For optically thin lines, such as Si~\textsc{iv} $1394\,\mathrm{\AA}$, double Gaussian fitting could give two averaged LOS velocities. For optically thick lines, such as Mg~\textsc{ii} k/h and H~\textsc{i} Lyman series, the profiles are often centrally reversed due to self-absorption, and effects of non-LTE (departure from the local thermodynamic equilibrium) should be considered \citep{2006A&A...459..651S,2015A&A...577A..92S,2008A&A...490..307G,2010A&A...514A..43G}. Despite these difficulties, we could still obtain some information of LOS velocities from the optically thick lines. \citet{2013ApJ...772...90L} proposed that for reversed profiles of Mg~\textsc{ii} k/h lines, the Doppler shift of the central minimum (or the maximum for a purely emission profile, called $\mathrm{k_3}$ for the k line, see Figure~\ref{fig:iris_cog}(j)) correlates strongly with the LOS velocity at the $\tau = 1$ height of the line core ($\tau$ represents the optical thickness), and the average Doppler shift of the peaks (called $\mathrm{k_{2b}}$ and $\mathrm{k_{2r}}$ for the blue and red sides of the peaks, respectively) correlates with the LOS velocity at the average $\tau = 1$ height of the peaks (deeper than the formation height of the line core). LOS velocity also affects the asymmetry of the peaks \citep{2008A&A...490..307G}. However, it is often difficult to identify the positions of $\mathrm{k_{2b}}$, $\mathrm{k_{2r}}$, and $\mathrm{k_3}$ for complex profiles.

The center of gravity (COG) method can be used to derive the Doppler shifts of purely absorption \citep[photospheric lines, ][]{2003ApJ...592.1225U} or purely emission lines. When there are multi-velocity components in a spectral profile, the COG method gives a weighted average result. However, the COG method is physically wrong for the emission profiles with central reversals, although this method is sometimes still useful in such case \citep{2019A&A...624A..72Z}. The COG method is expressed as
\begin{equation} \label{eq:moment}
     v_\mathrm{D} = \frac{\displaystyle \sum_i (I(v_i)-C)v_i}{\displaystyle \sum_i (I(v_i)-C)} \, ,
\end{equation}
where $v_\mathrm{D}$ is Doppler velocity, $I$ is intensity, $C$ is continuum, and wavelength $\lambda$ is converted into velocity ($v$) in the unit of ``$\mathrm{km\,s^{-1}}$'' using 
\begin{equation}
    v_i = \frac{\lambda_i-\lambda_0}{\lambda_0}c \, ,
\end{equation}
in which $\lambda_0$ is the rest wavelength and $c$ is the light speed. 

\begin{table}{h}
    \centering
    \caption{\label{tab:cog} Estimation of LOS velocities of the eight profiles shown in Figures~\ref{fig:iris_cog}(d)-(k) using different methods: shift of $\mathrm{k_3}$, average shift of $\mathrm{k_{2b}}$ and $\mathrm{k_{2r}}$, average shift at half maximum, and COG. The velocities are in the unit of ``$\mathrm{km\,s^{-1}}$''.}
    ~\\
    \begin{threeparttable}
    \begin{tabular}{r|cccccccc} 
    \hline
    Method & 56a & 56b & 56c & 56d & 62a & 62b & 62c & 62d \\
    \hline
    $\mathrm{k_3}$ & - & - & 16 & - & 16 & - & -5 & 6 \\
    $\mathrm{k_2}$ & - & 6 & - & - & - & - & 3 & 11 \\
    Half maximum & -6 & 6 & 12 & 11 & 16 & -9 & 3 & 8 \\
    COG & -7.1 & 6.9 & 11.3 & 13.4 & -11.6 & -10.8 & 3.6 & 6.1 \\
    \hline
    \end{tabular}
        \footnotesize
        NOTE--The given accuracies of the calculated velocities using the shift of $\mathrm{k_3}$, average shift of $\mathrm{k_2}$, and average shift at half maximum are $1\,\mathrm{km\,s^{-1}}$ due to that the spectral resolution is $\approx 5.5\,\mathrm{km\,s^{-1}}$, and the accuracies using the COG method are $0.1\,\mathrm{km\,s^{-1}}$ because around 40 pixels are considered.
\end{threeparttable}
\end{table}

To check the reliability of the COG method, we compare it with the methods proposed by \citet{2013ApJ...772...90L}. Figure~\ref{fig:iris_cog}(a) is a slit-jaw $2796\,\mathrm{\AA}$ image during the first scan, and Figures~\ref{fig:iris_cog}(b)-(c) show the images of the Mg~\textsc{ii} k line spectra along slit positions No.56 (at 11:07:39~UT) and No.62 (at 11:08:44~UT), respectively. Four points along the slit position No.56 and four along the No.62 are chosen (denoted with ``a'', ``b'', ``c'', ``d'' for different heights in panel (a)), and their Mg~\textsc{ii} k line profiles are plotted in Figures~\ref{fig:iris_cog}(d)-(k) sequentially. Positions of $\mathrm{k_3}$ (marked with the red plus symbols) and $\mathrm{k_2}$ (marked with the green stars) are identified when there is no much confusion. We also calculate the average shift at half maximum for each profile (marked with the blue $\times$ symbols). The results are listed in Table~\ref{tab:cog}. From the images of the spectra in Figures~\ref{fig:iris_cog}(b)-(c), it is intuitive that the top part along the slit position No.56 is slightly blue-shifted and the lower part of the prominence is red-shifted; along the slit position No.62, the top part has multi-velocity components and the lower part is also red-shifted. Among the eight line profiles in Figures~\ref{fig:iris_cog}(d)-(k), the positions of $\mathrm{k_3}$ are only identified in four profiles, and three of them are even questionable (56c, 62c, and 62d). So the LOS velocities derived from $\mathrm{k_3}$ are also questionable. The positions of $\mathrm{k_{2b}}$ and $\mathrm{k_{2r}}$ are determined in three profiles relatively precisely. The average shift at half maximum is calculated for each profile, and the results are consistent with what we see in the image of the spectra. However, this method may not include weak components as in case of the profiles from Positions 56a and 62a (Figures~\ref{fig:iris_cog}(d) and (h)). The LOS velocities derived from the COG method have the same signs as those derived from the average shifts of peaks or at half maximum except the 62a profile due to the aforementioned reason. The results of the COG method are also consistent with the images of the spectra, which means that this method is not influenced by occurrence of central reversals for the chosen profiles.

We will see that some Mg~\textsc{ii} k line profiles are reversed deeply during the prominence eruption. In most cases, the signs of LOS velocities derived from the COG method are generally consistent with the images of the spectra. We will give quantitative results using Gaussian fitting or directly from the images of the spectra, and use the COG method to derive LOS velocities statistically (not physically for profiles with self-absorption) and qualitatively.


\section{Results} \label{sec:res}
\subsection{Eruption overview}
The long filament, when the prominence is seen against the solar disk, extends from nearly solar disk center to beyond the northeastern solar limb on 2015 April 28th. Figures~\ref{fig:aia_evo}(a)-(e) show its EUV images, including the eruption process in AIA $304\,\mathrm{\AA}$. When seen in AIA 171 and $193\,\mathrm{\AA}$ (Figures~\ref{fig:aia_evo}(d)-(e)), the prominence is thinner than that seen in AIA $304\,\mathrm{\AA}$ (Figure~\ref{fig:aia_evo}(a)) due to lower opacities. A part of the prominence is blocked by itself before the prominence eruption (Figure~\ref{fig:aia_evo}(a)), and LOS is mainly along the prominence axis for IRIS observations. In Figure~\ref{fig:aia_evo}(b), the prominence is erupting; a dark region along the filament spine and two bright ribbons on the solar disk can be seen. In Figure~\ref{fig:aia_evo}(c), the prominence erupts further, and its spine inclines toward the solar equator. We synthesize a time-distance diagram in AIA $304\,\mathrm{\AA}$ along the slice AB in Figure~\ref{fig:aia_evo}(a), and the result is shown in Figure~\ref{fig:aia_evo}(f). We fit the prominence height in Figure~\ref{fig:aia_evo}(f) with a slow-rise phase (linear function) and a fast-rise phase (exponential function) using the approximation \citep{2013ApJ...769L..25C, 2015ApJ...807..144S}
\begin{equation}
    h(t) = c_0e^{(t-t_0)/\tau}+c_1(t-t_0)+c_2 \, ,
\end{equation}
where $h$ represents height, $t$ is time, $t_0$ is arbitrary, and $\tau$, $c_0$, $c_1$, $c_2$ are parameters obtained by fitting. The obtained initial rise speed is $c_1 = 3.78\,\mathrm{km\,s^{-1}}$, and the erupting speed at the end of our tracking is $\sim 103\,\mathrm{km\,s^{-1}}$. Onset of the fast-rise phase is defined by
\begin{equation}
    t_\mathrm{onset} = \tau \mathrm{ln}\left(\frac{c_1 \tau}{c_0}\right)+t_0 \, ,
\end{equation}
which is calculated to be 13:25:30~UT (the cyan vertical line in Figure~\ref{fig:aia_evo}(f)).

\subsection{Spectral evolution}
\begin{figure}[h!]
    \begin{center}
    \includegraphics[width=180mm]{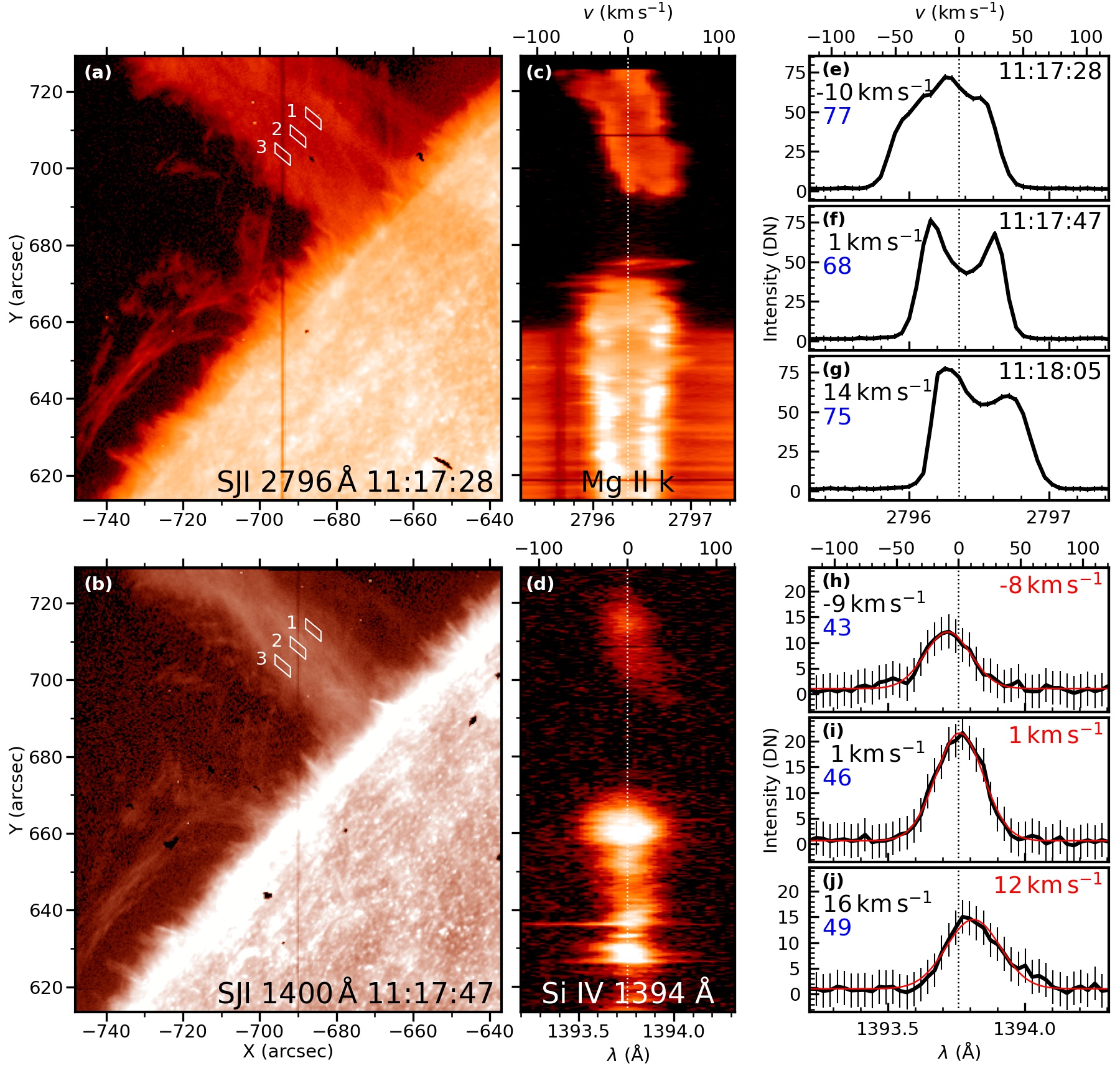}
    \end{center}
    \caption{\label{fig:spe1117} Spectral observations around 11:17~UT. First column: IRIS SJIs. Second column: images of the spectra at the slit position as shown in panel (a). Third column: average spectral profiles from boxes 1-3 in panels (a)-(b), sequentially. The first row is for Mg~\textsc{ii} observations and the second row is for the Si~\textsc{iv} window. The vertical dotted lines in (c)-(j) mark the positions of rest wavelengths. In panels (e)-(j), Doppler velocities derived from the COG method (black fonts) and FWHMs in the units of ``$\mathrm{km\,s^{-1}}$'' (blue fonts) are noted in left. In panels (h)-(j), the Doppler velocities derived from Gaussian fitting (red curves) are marked in right (red fonts).}
\end{figure}

\begin{figure}[h!]
\begin{center}
\includegraphics[width=180mm]{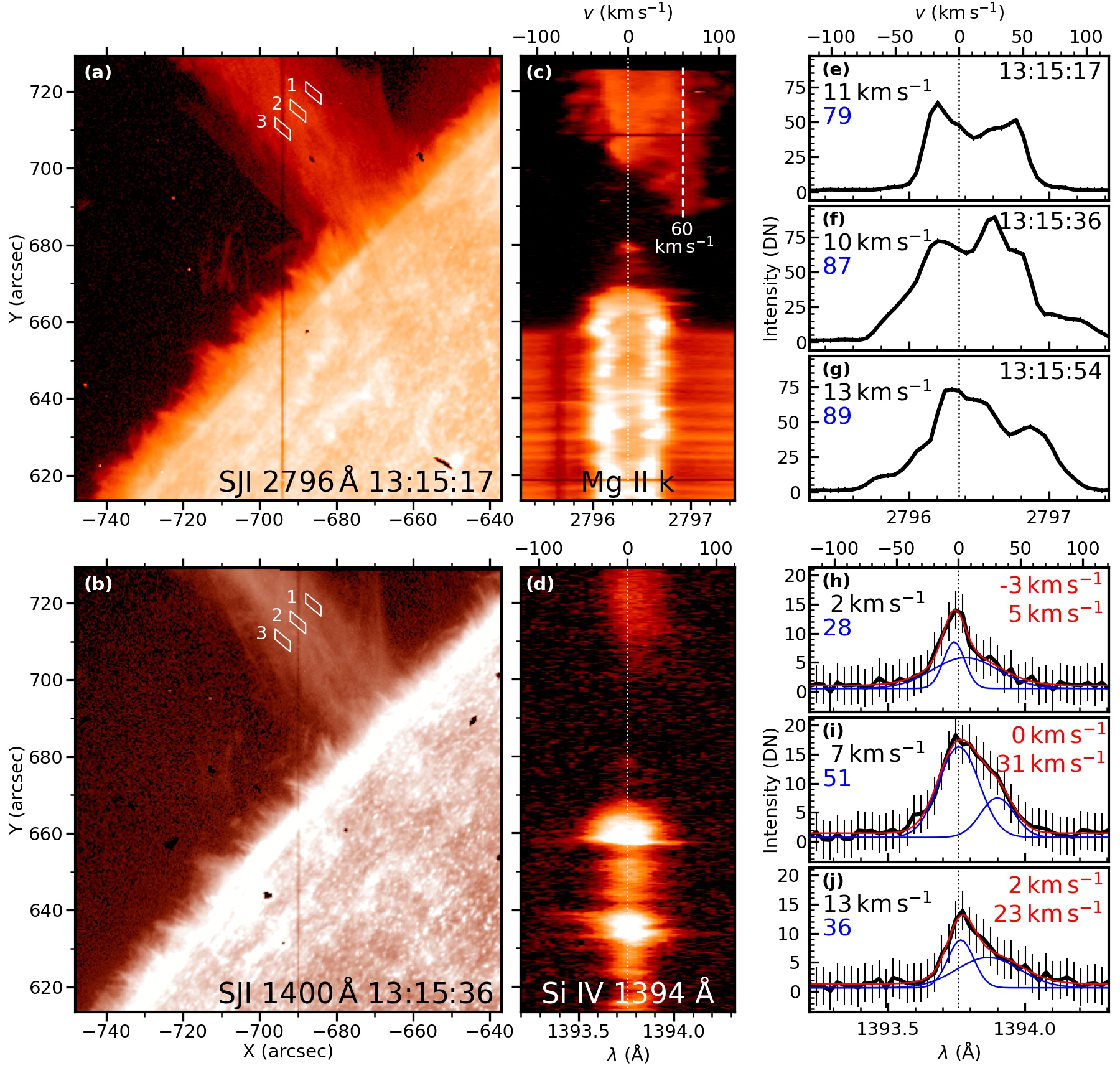}
\end{center}
\caption{\label{fig:spe1315} Spectral observations around 13:15~UT. The layout is similar to figure~\ref{fig:spe1117} but double Gaussian fitting is used for Si~\textsc{iv} $1394\,\mathrm{\AA}$ line profiles in panels (h)-(j), where blue curves show separate Gaussian components.}
\end{figure}

\begin{figure}[h!]
\begin{center}
\includegraphics[width=180mm]{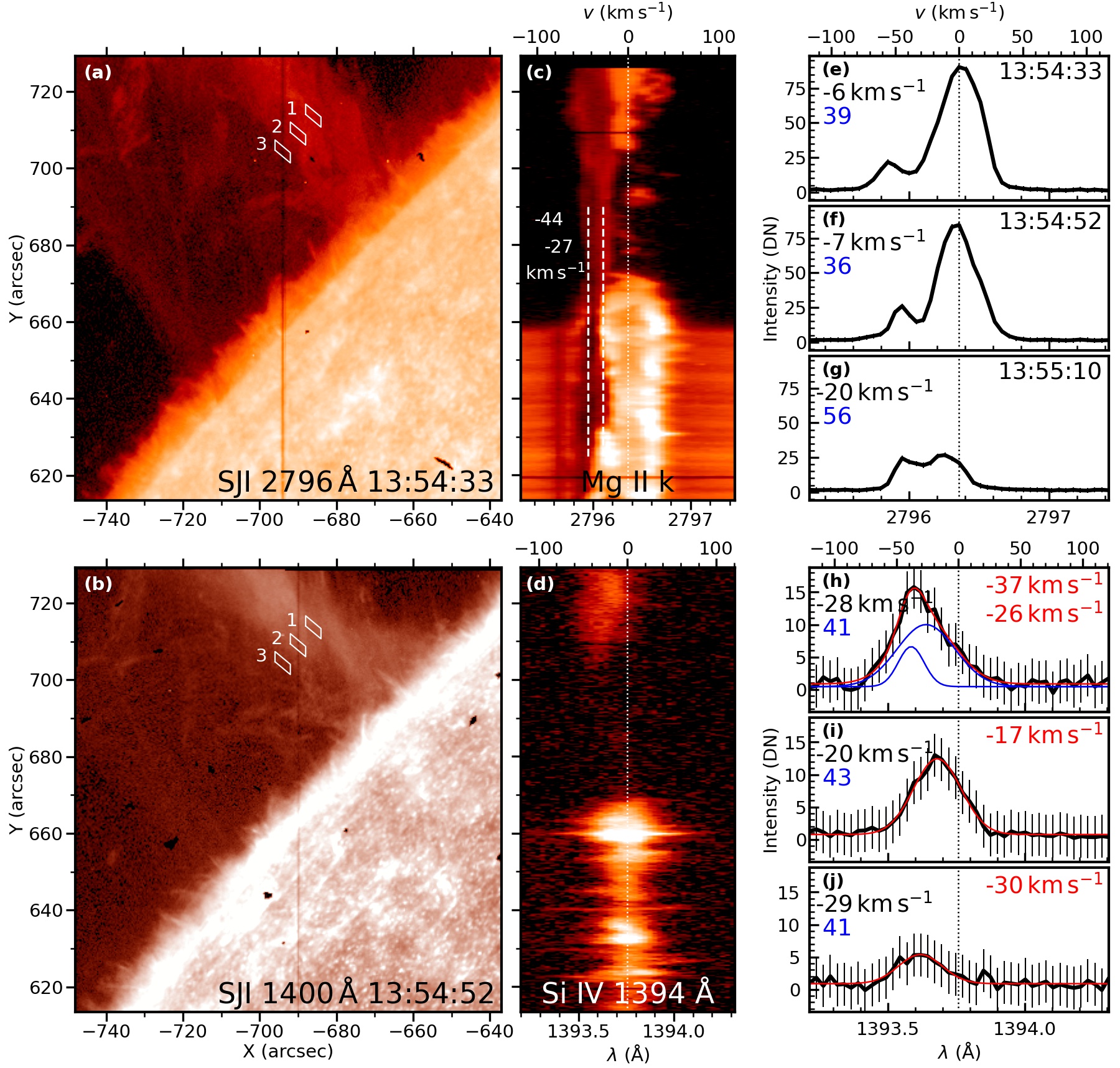}
\end{center}
\caption{\label{fig:spe1354} Spectral observations around 13:54~UT, similar to figures~\ref{fig:spe1117} and \ref{fig:spe1315}.}
\end{figure}

\begin{figure}[h!]
    \begin{center}
    \includegraphics[width=7in]{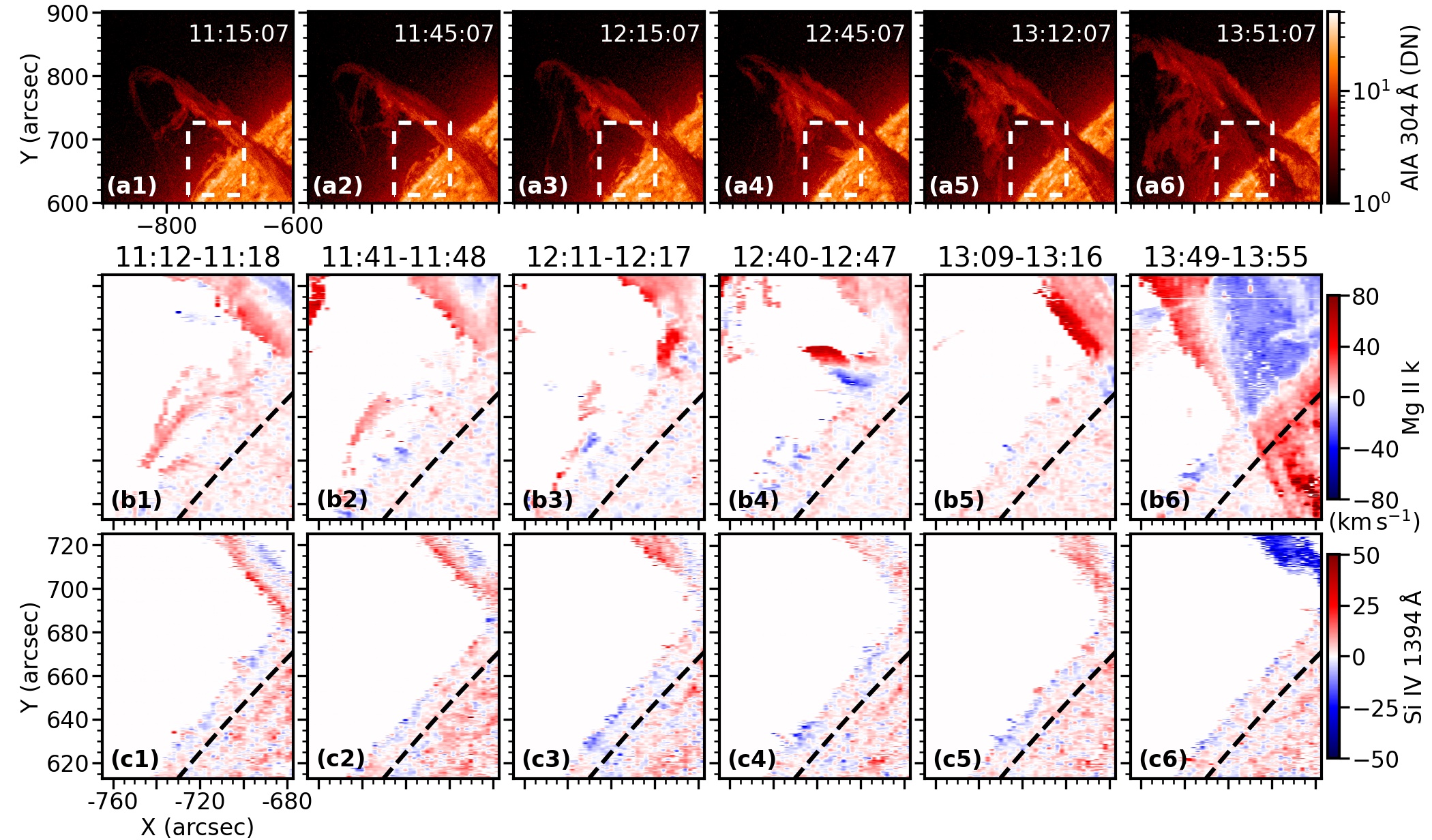}
    \end{center}
    \caption{\label{fig:dimage} Doppler images derived from the spectra of Mg~\textsc{ii} k (first row) and Si~\textsc{iv} $1394\,\mathrm{\AA}$ (second row) using the COG method. First row: the AIA $304\,\mathrm{\AA}$ images. Second and third rows: Doppler images from the Mg~\textsc{ii} k and the Si~\textsc{iv} $1394\,\mathrm{\AA}$ lines, respectively. Each Doppler image is composed of 44 raster steps from the slit position No.21 (left) to the slit position No.64 (right), and the observation times of the two slits are denoted above. The dashed black lines represent the solar limb. Color bars are shown at the right for each row.}
\end{figure}
We have seen in Figure~\ref{fig:iris_cog} around 11:07~UT that the top part of the raster is slightly blue-shifted and the lower part of the prominence is red-shifted. In this section, we will introduce the spectral evolution, especially the variations of LOS velocities of the erupting prominence in detail. Figure~\ref{fig:spe1117} shows IRIS observations around 11:17~UT (the second raster scan), including SJI snapshots (Figure~\ref{fig:spe1117} left column), images of the spectra along the dark slit in Figure~\ref{fig:spe1117}(a) (middle), and spectral profiles (right) of the Mg~\textsc{ii} k line (upper part) and the Si~\textsc{iv} line (lower part), respectively. Comparing the prominence images in the two wavelengths, SJI $1400\,\mathrm{\AA}$ has a narrower bright core with faint prominence edges (Figure~\ref{fig:spe1117}(b)) than SJI $2796\,\mathrm{\AA}$ (Figure~\ref{fig:spe1117}(a)), and the Si~\textsc{iv} $1394\,\mathrm{\AA}$ line profiles (Figures~\ref{fig:spe1117}(h)-(j)) are weaker with lower signal-to-noise ratios (S/N) than the Mg~\textsc{ii} k line (Figures~\ref{fig:spe1117}(e)-(g)). Mg~\textsc{ii} k line profiles are sometimes reversed (Figures~\ref{fig:spe1117}(f)-(g)), while Si~\textsc{iv} $1394\,\mathrm{\AA}$ line profiles can be fitted using a Gaussian profile (red curves in Figures~\ref{fig:spe1117}(h)-(j)). Besides, the Mg~\textsc{ii} k line has a larger line width than the Si~\textsc{iv} $1394\,\mathrm{\AA}$, despite that the latter is formed at a higher temperature. These performances are mainly due to that the Mg~\textsc{ii} k line has a larger opacity than the Si~\textsc{iv} $1394\,\mathrm{\AA}$.

In both of the images of the Mg~\textsc{ii} k and Si~\textsc{iv} $1394\,\mathrm{\AA}$ spectra in Figures~\ref{fig:spe1117}(c)-(d), the blue-shifted top part and red-shifted lower part are visible. We choose 3 regions in Figures~\ref{fig:spe1117}(a)-(b) from right to left with respect to the prominence axis, and their average spectral profiles are plotted in Figures~\ref{fig:spe1117}(e)-(j). In these panels, the Doppler velocities derived from the COG method (black fonts) and the full widths at half maximum (FWHMs, blue fonts, in the unit of ``$\mathrm{km\,s^{-1}}$'') are denoted at the left. In Figures~\ref{fig:spe1117}(h)-(j) for the Si~\textsc{iv} $1394\,\mathrm{\AA}$ line profiles, the Doppler velocities from Gaussian fitting are denoted at the right (red fonts). Both Mg~\textsc{ii} k and Si~\textsc{iv} $1394\,\mathrm{\AA}$ lines show that the right part of the prominence is blue-shifted, with the Doppler velocity of $-8.4\pm 0.7\,\mathrm{km\,s^{-1}}$ (Figure~\ref{fig:spe1117}(h) using single Gaussian fitting); the left part is red-shifted, with the Doppler velocity of $11.7\pm 0.8\,\mathrm{km\,s^{-1}}$ (Figure~\ref{fig:spe1117}(j)). 

The prominence spectra vary obviously when the prominence approaches eruption. The layout of Figure~\ref{fig:spe1315} is similar to Figure~\ref{fig:spe1117} but observed at $\sim$13:15~UT, about 10 minutes before the onset of the fast-rise phase. In Figure~\ref{fig:spe1315}(c), a red-shifted (around $60\,\mathrm{km\,s^{-1}}$) faint component appears, which can be seen in SJI at the prominence edges (Figures~\ref{fig:spe1315}(a)-(b)). The Mg~\textsc{ii} k line profiles in Figures~\ref{fig:spe1315}(e)-(g) are red-asymmetry with multi-velocity components, and another red Gaussian profile is necessary to fit the Si~\textsc{iv} $1394\,\mathrm{\AA}$ line profiles in Figures~\ref{fig:spe1315}(h)-(j). With more velocity components, the line widths get wider, and FWHM of the Mg~\textsc{ii} k line is nearly $90\,\mathrm{km\,s^{-1}}$ (Figures~\ref{fig:spe1315}(f)-(g)), and that of the Si~\textsc{iv} $1394\,\mathrm{\AA}$ line is around $50\,\mathrm{km\,s^{-1}}$ (Figure~\ref{fig:spe1315}(i)). However, the Mg~\textsc{ii} k line of the faint component has a narrower width around $30\,\mathrm{km\,s^{-1}}$ (Figure~\ref{fig:spe1315}(c)).

During the fast-rise phase in Figure~\ref{fig:spe1354}, the faint components changes to be blue-shifted, with 2 main Doppler velocities shown in the image of the Mg~\textsc{ii} k line spectra (Figure~\ref{fig:spe1354}(c)): $-44$ and $-27\,\mathrm{km\,s^{-1}}$. In the image of Si~\textsc{iv} $1394\,\mathrm{\AA}$ spectra in Figure~\ref{fig:spe1354}(d), the bright component is also obviously blue-shifted, and the Gaussian fitting results show that the Doppler velocity is at least $-17\,\mathrm{km\,s^{-1}}$ (Figure~\ref{fig:spe1354}(i)). However, the bright component in Mg~\textsc{ii} k has no obvious shift (Figure~\ref{fig:spe1354}(c)), and the line widths are small (Figures~\ref{fig:spe1354}(e)-(f)). We will see in next section that it is due to the absorption by the faint component. The emission of the box 3 in the SJI $2796\,\mathrm{\AA}$ (Figure~\ref{fig:spe1354}(a)) is mainly from the faint component, which is identified from the weak emission of the Mg~\textsc{ii} k line with blue shifts (Figures~\ref{fig:spe1354}(g)). The weak emission of the Si~\textsc{iv} $1394\,\mathrm{\AA}$ line from the box 3 (Figures~\ref{fig:spe1354}(j)) with the similar blue shift suggests that the Si~\textsc{iv} line has a contribution to the brightness of the faint component in SJI $1400\,\mathrm{\AA}$.

To analyze the evolution of the prominence LOS velocities in detail, we calculate Doppler images from both the Mg~\textsc{ii} k and Si~\textsc{iv} $1394\,\mathrm{\AA}$ lines using the COG method. After the calculations, isolated noises in the Doppler images are further removed. 

The obtained Doppler images are shown in Figure~\ref{fig:dimage}, where the first row shows AIA $304\,\mathrm{\AA}$ images and the dashed boxes mark the FOV of the Doppler images in the lower two rows. Initially (the first column in Figure~\ref{fig:dimage}), the left part of the prominence is red-shifted and the right part is mainly blue-shifted with a boundary near the prominence axis; the maximum red-shifted velocity is $\sim 30\,\mathrm{km\,s^{-1}}$, and the maximum blue-shifted velocity is $\sim 20\,\mathrm{km\,s^{-1}}$ from both Mg~\textsc{ii} k and Si~\textsc{iv} $1394\,\mathrm{\AA}$ Doppler images (the difference is $<2\,\mathrm{km\,s^{-1}}$). Then, red shifts dominate the prominence gradually, and the largest Doppler shifts always occur at the left edge. At around 13:15 (Figure~\ref{fig:dimage}(b5)), there is a largely red-shifted region corresponding with the faint component as shown in Figures~\ref{fig:spe1315}(a)-(b). About 25 minutes after the onset of the fast-rise phase (the rightmost column in Figure~\ref{fig:dimage}), the erupting prominence is mainly blue-shifted, and the average and maximum LOS velocities in Si~\textsc{iv} $1394\,\mathrm{\AA}$ window are 22 and $47\,\mathrm{km\,s^{-1}}$, respectively. The blue-shifted velocities in Mg~\textsc{ii} k window are smaller, and red shifts can still be seen at the left edge. Note that in the Mg~\textsc{ii} k Doppler image in Figure~\ref{fig:dimage}(b6), the positive values along the filament on the solar disk result from the fact that part of the blue wing of the Mg~\textsc{ii} k line, which is radiated from the solar disc, is absorbed by the erupting filament (see Figure~\ref{fig:spe1354}(c)).

\subsection{Faint component}
\begin{figure}[h!]
    \begin{center}
    \includegraphics[width=180mm]{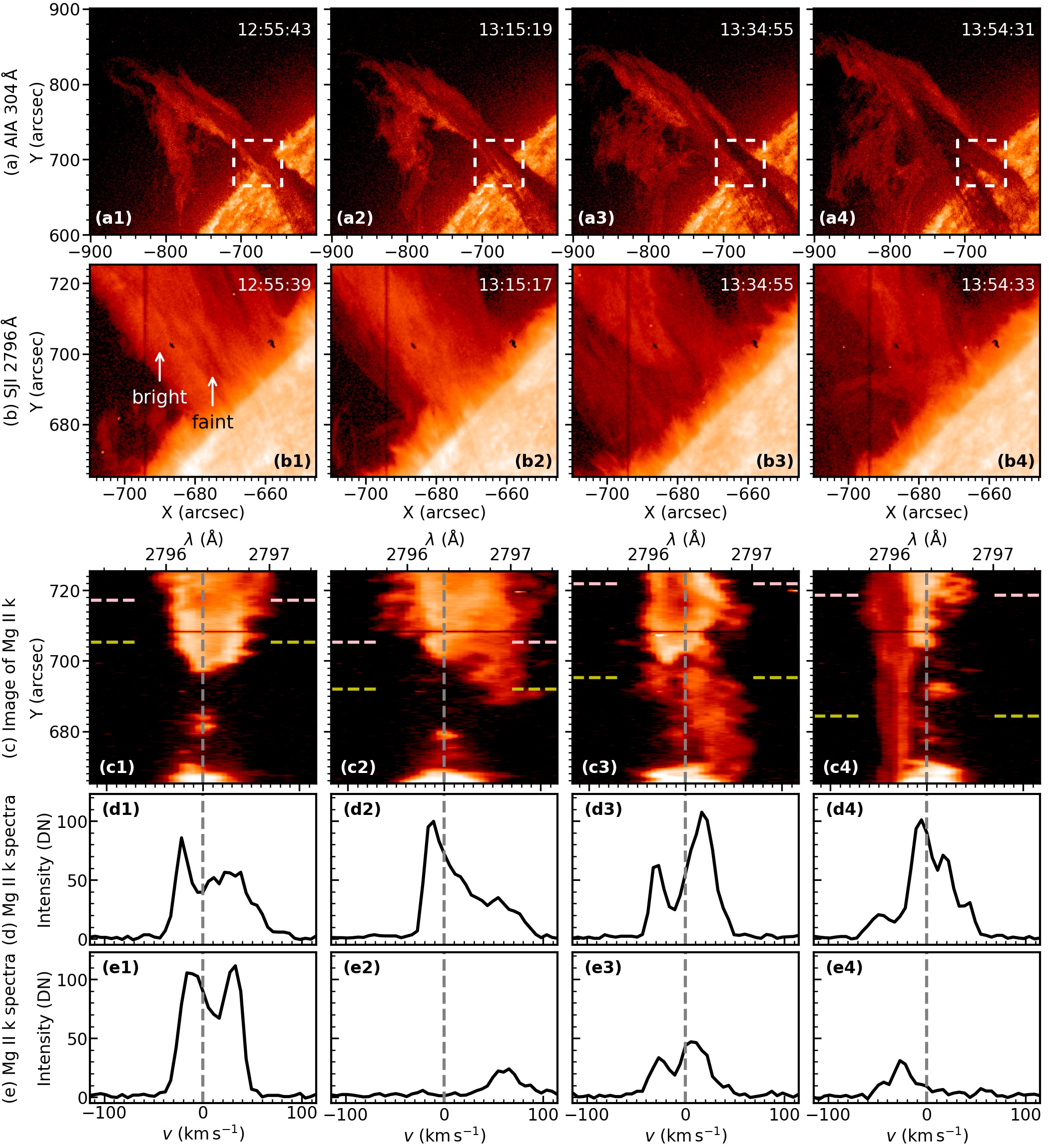}
    \end{center}
    \caption{\label{fig:faint} AIA $304\,\mathrm{\AA}$ and Mg~\textsc{ii} k observations focusing on faint components. (a) The AIA $304\,\mathrm{\AA}$ images. The dashed boxes represent the FOV of the snapshots in panels (b). (b) The slit-jaw $2796\,\mathrm{\AA}$ images. (c) Images of the Mg~\textsc{ii} k spectra at the slit positions as shown in images in panels (b). (d) Mg~\textsc{ii} k profiles from the positions marked with dashed pink lines in panels (c). (e) Mg~\textsc{ii} k profiles from the positions marked with dashed yellow lines in panels (c).}
\end{figure}
A detailed view on the evolution of the faint component is shown in Figure~\ref{fig:faint}, where the AIA $304\,\mathrm{\AA}$ images at different times are shown in the first row, the slit-jaw $2796\,\mathrm{\AA}$ images are shown in the second row, the third row shows images of the Mg~\textsc{ii} k line spectra along the slit positions in Figure~\ref{fig:faint}(b), and the remaining two rows show Mg~\textsc{ii} k profiles at the positions marked in Figure~\ref{fig:faint}(c). Panels (c)-(e) share the same abscissa but in different units. At 12:55:39~UT (the first column in Figure~\ref{fig:faint}), the regions with or without the faint components can be differed in both AIA $304\,\mathrm{\AA}$ and SJI $2796\,\mathrm{\AA}$. In Figure~\ref{fig:faint}(c1), the lower part of prominence spectrum is not blended by the faint component, which is reversed with 2 clear peaks (Figure~\ref{fig:faint}(e1)); but the upper part is overlaid with the faint component and the peak at the red wing is absorbed with some weak peaks (Figure~\ref{fig:faint}(d1)). Twenty minutes later (the second column in Figure~\ref{fig:faint}, the same time as Figure~\ref{fig:spe1315}), the faint component is significantly red-shifted. From Figure~\ref{fig:faint}(c2), we can see that the largest Doppler velocity of red-shifted component is beyond $100\,\mathrm{km\,s^{-1}}$. The Mg~\textsc{ii} k line in Figure~\ref{fig:faint}(d2) includes information of both bright and faint components; it also peaks at blue wing and the red wing is absorbed. At 13:34:55~UT, about 10 minutes after the onset of fast-rise phase (the third column in Figure~\ref{fig:faint}), the sign of Doppler shifts of the faint component is changing: the lower part is mainly red-shifted but the upper part is blue-shifted (Figure~\ref{fig:faint}(c3)). The Mg~\textsc{ii} k line in Figure~\ref{fig:faint}(d3) is seriously reversed, and the ratio of line peak to central minimum is $4.5$. As a comparison, the ratio of the profile in Figure~\ref{fig:faint}(e1) is $1.7$. At 13:54:33~UT (the fourth column in Figure~\ref{fig:faint}, the same time as Figure~\ref{fig:spe1354}), the faint component is blue-shifted, and the blue wing of the bright component is absorbed. When focusing on the AIA $304\,\mathrm{\AA}$ images, we can see the darkening of the faint region during the eruption process. The expansion of the faint prominence spine is visible in both AIA $304\,\mathrm{\AA}$ and SJI $2796\,\mathrm{\AA}$.

\section{Discussion} \label{sec:dis}
The prominence eruption in the FOV of SDO/AIA can be divided into slow- and fast-rise phases, and the onset of the fast-rise phase is around 13:25~UT. However, The prominence Doppler shifts experience three periods: (1) two hours before the fast-rise phase, the left part of prominence, with respect to the prominence axis, is red-shifted and the right part is mainly blue-shifted (first column in Figure~\ref{fig:dimage}); (2) then, red-shifted area increases, and almost the whole prominence (in IRIS FOV) is red-shifted at 10 minutes before the fast-rise phase; (3) during the fast-rise phase, the prominence gets to be blue-shifted. A faint component is clearly identified at 12:55:39~UT (first column in Figure~\ref{fig:faint}). The faint component in Mg~\textsc{ii} k window has a narrow line width ($\sim 30\,\mathrm{km\,s^{-1}}$, Figure~\ref{fig:spe1315}) and significant variations of Doppler shifts. Besides, the faint component can absorb the radiation from the bright part, which leaves a single profile peak when the faint component has a significant Doppler shift, or results in a deep central reversal when the shift is slight. In Si~\textsc{iv} window, the faint component can still be identified in SJI despite the lower S/N; two Gaussian profiles are necessary to fit the Si~\textsc{iv} $1394\,\mathrm{\AA}$ line profiles when the bright component is overlapped with the faint component in LOS. The faint region along the prominence spine in AIA $304\,\mathrm{\AA}$ images is spatially related to the faint component. The darkening and expansion of the faint region during the prominence eruption are visible. The faint component is also red-shifted before the fast-rise phase, with a larger LOS velocity than the bright component. In the fast-rise phase, the faint component gets to be blue-shifted, too. In the following sections, we will give our explanations on these phenomena.

\subsection{Magnetic configuration and view of the prominence}
\begin{figure}[h!]
    \begin{center}
    \includegraphics[width=180mm]{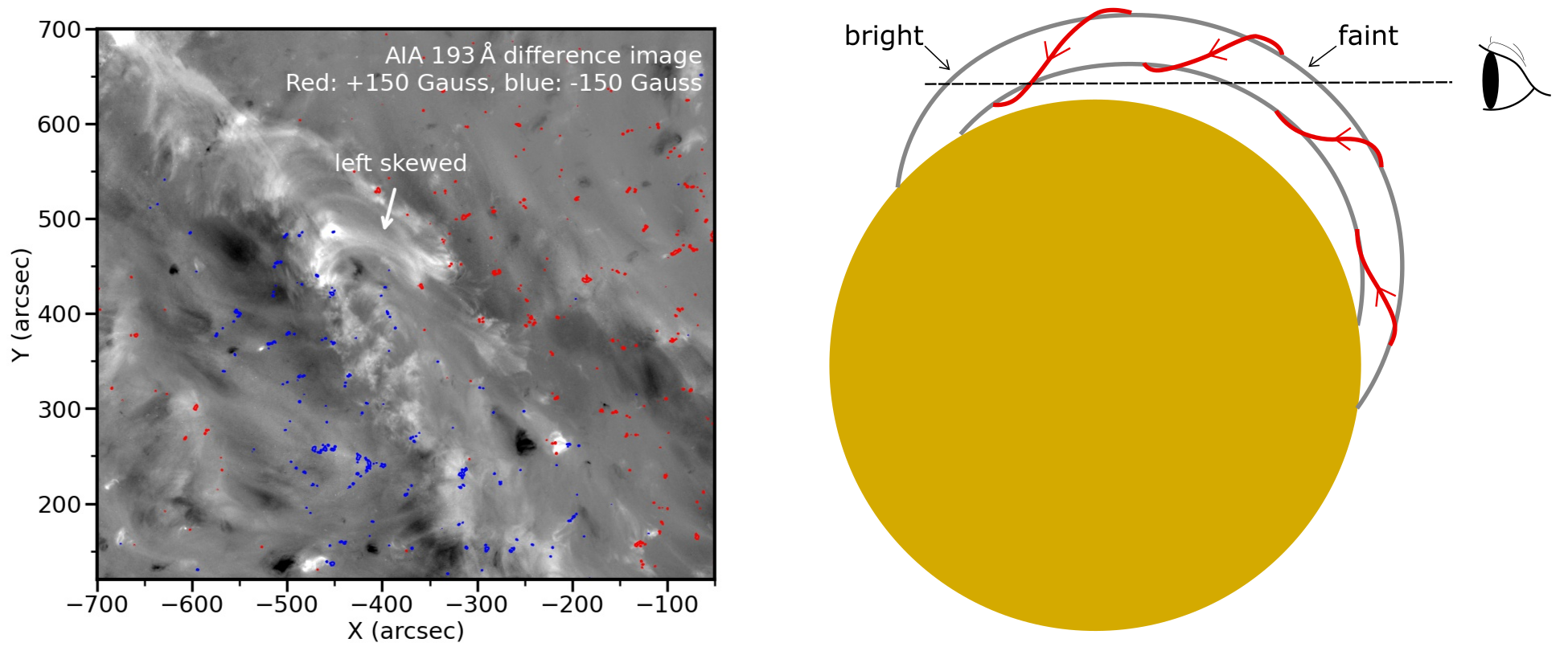}
    \end{center}
    \caption{\label{fig:cartoon} Prominence chirality and view. (a) AIA $193\,\mathrm{\AA}$ difference image by the image observed at 16:00:06~UT subtracting the one at 11:57:06~UT. The red and blue contours mark LOS magnetic field $\pm 150$ Gauss, respectively, from HMI observed at 15:59:05~UT. (b) Cartoon showing the prominence viewing angle. The red arrows represent the magnetic helicity with exaggerated twist number.}
\end{figure}

\begin{figure}[h!]
    \begin{center} 
    \includegraphics[width=85mm]{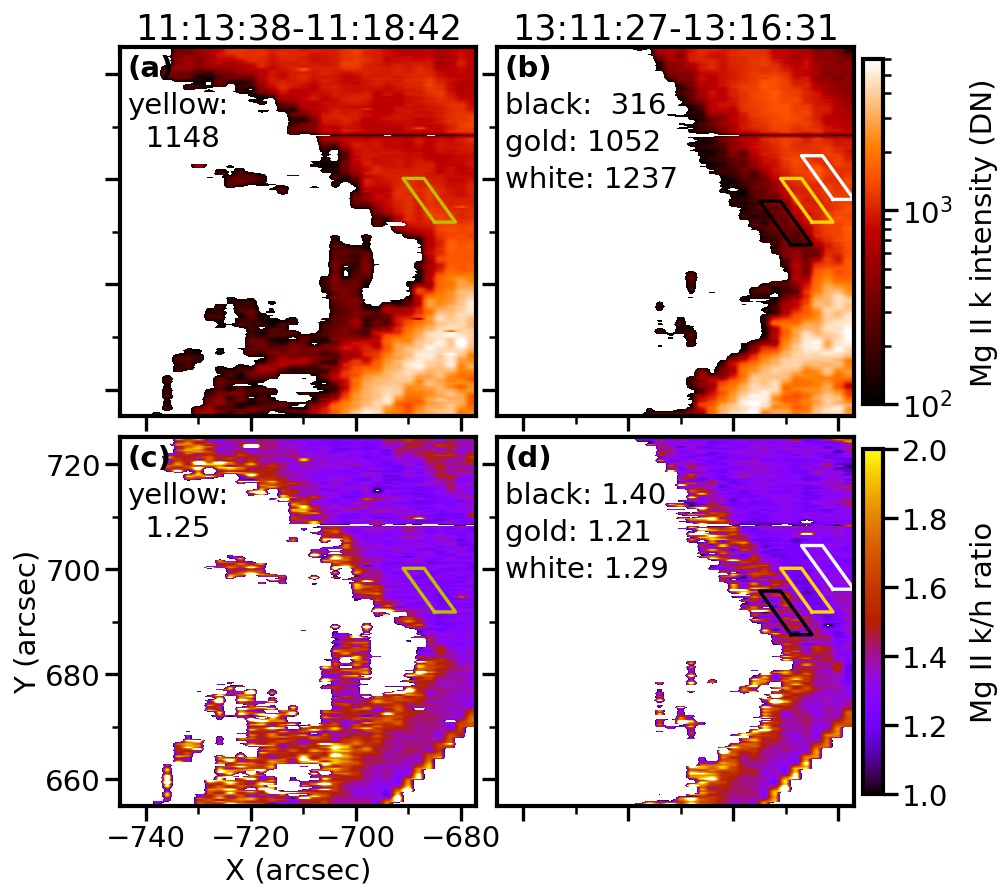}
    \end{center}
    \caption{\label{fig:khratio} Images of integrated uncalibrated intensity of the Mg~\textsc{ii} k line (a)-(b) and the Mg~\textsc{ii} k/h intensity ratio (c)-(d). Mean values of the marked regions are noted in each panel.}
\end{figure}
A knowledge about the magnetic configuration and viewing angle of the prominence is helpful for understanding its observational features. The left image in Figure~\ref{fig:cartoon} is a AIA $193\,\mathrm{\AA}$ difference image overlaid with LOS magnetogram contours of $\pm 150\,\mathrm{Gauss}$. The post-eruption loops is left skewed, which suggests that the filament channel has a negative magnetic helicity \citep{2014ApJ...784...50C}. Hence, we plot the cartoon of the filament as a flux rope \citep{2017ApJ...835...94O} with an exaggerated twist number for a clear view (the right image in Figure~\ref{fig:cartoon}). Due to the fact that the faint component can absorb Mg~\textsc{ii} k radiation from the bright component, the faint component should be in front, which could also be seen in AIA $304\,\mathrm{\AA}$ images (Figure~\ref{fig:faint}(a)). 

\subsection{Causes of Doppler evolutions}

Opposite Doppler shifts observed in a same prominence are generally explained as counter-streaming. In previous observations of counter-streaming, bidirectional flows were seen thread by thread \citep{1998Natur.396..440Z}, or blue- and red-wing images showed different directional flows \citep{2003SoPh..216..109L}. Some of counter-streaming observations could be explained by local motions of plasma, such as small-scale oscillations \citep{2010ApJ...721...74A} or magnetic reconnection \citep{2015ApJ...814L..17S}. In our event, the opposite Doppler shifts are found from the beginning of IRIS observations, and are possibly a property of the polar crown prominence that we studied. A major difference from previous observations is that the opposite Doppler shifts are distributed much regularly, i.e., the left half is red-shifted and the right part is mainly blue-shifted. This phenomenon reveals large-scale flows along the prominence spine, and cannot be explained by the local plasma motions.

Unidirectional flows along the flux rope could also cause opposite Doppler shifts when the LOS is perpendicular to the flux rope axis \citep{2014A&A...570A..93L,2019A&A...626A..98L,2014ApJ...784...50C,2017ApJ...836..122Z}. Despite that the LOS is mainly along the prominence spine in the FOV of the IRIS in our case, we assume that the opposite Doppler shifts are caused by the flows within the spine. In this case, in the cartoon of Figure~\ref{fig:cartoon}, the flows of the part marked with ``bright'' should move toward the observer, or the flows marked with ``faint'' should move away from the observer, or both, to cause the left being red-shifted and the right being blue-shifted with respect to the prominence axis. In the following period, the spine is dominantly red-shifted, but the blue shifts are not enhanced. The different evolutions of the red and blue shifts makes this assumption questionable.

During the fast-rise phase in the third column in Figure~\ref{fig:faint}, the faint component shows opposite Doppler shift along the slit, which is possibly caused by untwisting motion of the erupting prominence \citep{2015ApJ...803...85L}, which could be seen in AIA $304\,\mathrm{\AA}$ images by tracking the faint region (Figure~\ref{fig:faint}).

When approaching the fast-rise phase, red shifts dominate the prominence, and the faint component has a large LOS velocity. In the Mg~\textsc{ii} k window, the LOS velocity of the bright component is not clear due to the fact that its emission is absorbed partly by the faint component. But from the Si~\textsc{iv} window, the results of the double Gaussian fitting (Figures~\ref{fig:spe1315}(h)-(j)) suggest that the Doppler shift of the bright component is not obvious. The red shifts could be explained by the mass drainage during the elevation of the prominence spine, which could accelerate the prominence eruption in return \citep{2020ApJ...898...34F}. In our case, if the prominence mass moves toward the prominence footpoint behind the solar disk, the direction of LOS velocity is consistent with the red shifts (see the cartoon in Figure~\ref{fig:cartoon}). The large Doppler velocity of the faint component suggests that the mass near the spine center moves faster than the mass near the footpoint, which is possibly due to more significant elevation of the prominence spine center, and that the large density near the footpoint may slow down the flows.

During the fast-rise phase, the prominence gets to be blue-shifted, which reveals the overall movement of the erupting prominence away from the solar disk and toward the observer.

\subsection{Formation of the faint component}
\citet{2015ApJ...803...85L} did also observe a faint and narrow component with a large Doppler velocity up to $460\,\mathrm{km\,s^{-1}}$ in an erupting active region prominence. The authors proposed that the two components with different Doppler velocities suggest that the erupting material is distributed in a hollow cone shape. The faint component in our observations share most of features as reported by \citet{2015ApJ...803...85L}, despite that the Doppler velocity is relatively small. In addition, there are some phenomena only seen in our observations. Firstly, we could see the faint component in AIA $304\,\mathrm{\AA}$ images and SJIs directly, hence we observed darkening and expansion of the faint component during the prominence eruption (Figures~\ref{fig:faint}(a)-(b)). Besides, we observed that the faint component can absorb the Mg~\textsc{ii} k radiation from the bright component. On the basis of these phenomena, we suspect that the Mg~\textsc{ii} k line from the faint component has a lower opacity than that from the bright component, and the faint component is composed of low-density and cold plasma due to the expansion of the prominence spine.

The intensity ratio of Mg~\textsc{ii} k and h lines is widely used to check the opacity \citep{2013ApJ...772...90L,2015ApJ...803...85L,2016SoPh..291...67V,2019A&A...624A..72Z}: the ratio is generally in the range of 1 to 2, and approaches 2 under the optically thin assumption and gets small with the increase of opacity.  Figure~\ref{fig:khratio} shows the images of Mg~\textsc{ii} k integrated intensity (panels (a)-(b)) and k/h ratio (panels (c)-(d)) from spectral observations. Continuum intensity is subtracted when calculating the images. The left column in Figure~\ref{fig:khratio} is observed between 11:13~UT and 13:19~UT before the occurrence of the faint component, and the k/h ratio is relatively homogeneous. A region is selected in Figure~\ref{fig:khratio}(c) with yellow box, and the mean k/h ratio is 1.25. The right column in Figure~\ref{fig:khratio} is observed between 13:11~UT and 13:17~UT, and a faint region is seen at the prominence left edge (Figure~\ref{fig:khratio}(b)). The k/h ratio image in Figure~\ref{fig:khratio}(d) shows that the faint component has a larger k/h ratio (1.40) than the bright prominence (mainly between 1.2 and 1.3), which suggests that the faint component has a lower opacity. It should be noticed that the k/h ratios vary for different prominences with different viewing angles. \citet{2019A&A...624A..72Z} reported the k/h value around 1.4 in the main body of an eruptive prominence in quiet region, but in the erupting prominence studied by \citet{2015ApJ...803...85L}, the intensity ratio of the primary bright component varies from 1.4 to 1.9.


Prominence Mg~\textsc{ii} k radiation is partly from the emission of local plasma, and partly from the scattering of the chromospheric radiation \citep{2014A&A...564A.132H,2016SoPh..291...67V}. In our observations, the faint component can absorb or scatter the light from the bright part behind. Using non-LTE radiative transfer techniques, \citet{2013ApJ...772...90L} simulated emergent Mg~\textsc{ii} k and h lines in solar disk; the authors found that Mg~\textsc{ii} k core intensity is weak if the line core forms at a high position due to a low density and three-dimensional scattering, although temperature increases along height in the chromosphere. In our observations, the low emission (Figures~\ref{fig:faint}(e2)-(e4)) and deep central reversal (Figure~\ref{fig:faint}(d3)) of the faint component possibly result from the scattering of low-density plasma. In addition, line width is mainly determined by plasma temperature and micro-turbulence. The narrow Mg~\textsc{ii} k line profiles suggest that the faint component is mainly prominence core, whose temperature is lower than the PCTR. Therefore, we propose that the faint component consists of low-density and cold plasma, which appear due to the expansion of the prominence core during the prominence eruption. In this process, the flows move fast along the prominence spine and result in significant Doppler shifts. However, non-LTE modeling is necessary to give a strict explanation on the characteristics of the AIA $304\,\mathrm{\AA}$ and IRIS Mg~\textsc{ii} k observations.

\section{Conclusion} \label{sec:con}
We studied spectral evolution of an eruptive polar crown prominence using its IRIS observations in the Mg~\textsc{ii} and Si~\textsc{iv} lines and AIA EUV images. The main observational results of this work are listed in Table~\ref{tab:res}. The AIA observations suggest that the prominence experiences a slow- and fast-rise phase before it leaves the FOV of AIA. Simultaneously, the variation of Doppler shifts of the erupting prominence could be divided into three periods. In the first period, more than 2 hours before the onset of the fast-rise phase, opposite Doppler shifts at the two sides of the prominence axis are found with maximum LOS velocity between $20-30\,\mathrm{km\,s^{-1}}$. In the second period, around the onset of the fast-rise phase, the whole prominence gets to be red-shifted gradually. In the third period, the prominence is dominantly blue-shifted. The possible cause of the opposite Doppler shifts in the first period is large-scale counter-streaming, or unidirectional flows along the prominence spine (as a flux rope). More observations are necessary to determine which mechanism results in the opposite shifts, then reveal the mode of flows within the prominence spine. Besides, the opposite Doppler shifts of the faint component during the fast-rise phase may result from the untwisting motion of the erupting prominence. The obvious red shifts in the second period may reveal mass drainage along the prominence spine due to the elevation of the prominence, and the mass drainage might accelerate the prominence eruption in return. The blue shifts in the last period is likely to result from the eruption of the prominence toward the observer. 
\begin{center}
    \captionof{table}{\label{tab:res} Main observational results of this work.}
    ~\\
    \begin{tabular}{r|l} 
    \hline
    Prominence eruption & (1) Slow-rise phase \\
      & (2) Fast-rise phase \\
      \hline
    Evolution of spectra & (1) Opposite Doppler shifts with respect to the prominence axis in slow-rise phase \\
      & (2) Dominantly red-shifted around the onset of fast-rise phase \\
      & (3) Dominantly blue-shifted during prominence eruption \\
    \hline
    Faint component & Faint and narrow features in the Mg~\textsc{ii} k line \\
      & Large red shifts ($\sim 60\,\mathrm{km\,s^{-1}}$) around the onset of fast-rise phase \\
      & Darkening and expansion in AIA $304\,\mathrm{\AA}$ \\
    \hline
    \end{tabular}
\end{center}

During the second period, a faint component appears in AIA $304\,\mathrm{\AA}$, SJI 2796 and $1400\,\mathrm{\AA}$. The faint component has a narrow line profile, is initially red-shifted with a typical LOS velocity of $60\,\mathrm{km\,s^{-1}}$. The Mg~\textsc{ii} k/h ratio of the faint component ($\sim 1.40$) is larger than that of the bright component (between 1.2 and 1.3), which suggests that the faint component has a lower opacity. We also observed the darkening and expansion of the faint component in AIA $304\,\mathrm{\AA}$ images. On the basis of these characteristics, we propose that the faint component is composed of low-density and cold plasma due to the expansion of the prominence during eruption.

Hence we can relate the evolution of the spectra and the formation of the faint component to the prominence eruption. The opposite Doppler shifts are properties of the polar crown prominence that we studied. When the prominence approaches eruption, the prominence spine elevates and expands, and the acceleration of the mass drainage causes the obvious red shifts. Simultaneously, a faint region along the prominence spine forms and gets darker due to the decreasing of the plasma density during the spine expansion. Finally, the acceleration of the prominence eruption results in the blue shifts. Despite the consistence of above explanations, however, non-LTE radiative transfer simulations in future are necessary to interpret the observational characteristics of the Mg~\textsc{ii} k line and AIA $304\,\mathrm{\AA}$ images.

\section*{Conflict of Interest Statement}

The authors declare that the research was conducted in the absence of any commercial or financial relationships that could be construed as a potential conflict of interest.

\section*{Author Contributions}
Jianchao Xue processed the observational data, decided the main content of the paper, plotted figures, and wrote the manuscript. Hui Li selected the topic, joined discussion, and modified the manuscript. Yang Su joined discussion, offered ideas about the explanation on the observed phenomena, and modified the manuscript.

\section*{Funding}
This work is supported by NSFC grants (11427803, U1731241, U1631242, 11820101002) and by CAS Strategic Pioneer Program on Space Science, grant Nos. XDA15052200, XDA15320103, XDA15320300, and XDA15320301. Yang Su acknowledges the Jiangsu Double Innovation Plan.

\section*{Acknowledgments}
The authors thank the reviewers for their suggestions, some findings and ideas in this paper are proposed by them. Jianchao Xue thanks Jean-Claude Vial, Yingna Su, and Ping Zhang for discussion, thanks Ying Li and Hui Tian for a guidance on processing IRIS data. We thank the IRIS, AIA, and HMI teams for providing the data.

\bibliographystyle{frontiersinSCNS_ENG_HUMS} 
\bibliography{promerup}

\end{document}